\documentclass[11pt,a4]{article}
\usepackage[margin=1.2in]{geometry}
\usepackage{graphicx}
\usepackage{epsfig}
\usepackage{amssymb,amsfonts,amsmath,amsthm}
\usepackage{mathrsfs}
\usepackage{epstopdf}
\usepackage{hyperref,color}
\usepackage{natbib}
\usepackage{setspace}

\newtheorem{thm}{Theorem}
\newtheorem{defi}{Definition}

\pdfoutput=1

\begin{document}

\title{Are black holes about information?}
\author{Christian W\"uthrich\thanks{I wish to thank Karen Crowther, Erik Curiel, John Dougherty, Nick Huggett, Niels Linnemann, Keizo Matsubara, Tim Maudlin, John Norton, Daniele Oriti, Carlo Rovelli, and Karim Th\'ebault for discussions and comments, as well as audiences in Munich and D\"usseldorf. This work was performed under a collaborative agreement between the University of Illinois at Chicago and the University of Geneva and made possible by grant number 56314 from the John Templeton Foundation. Its contents are solely the responsibility of the author and do not necessarily represent the official views of the John Templeton Foundation.}}
\date{20 September 2017}
\maketitle

\begin{center}
In Richard Dawid, Radin Dardashti, and Karim Th\'ebault (eds.), {\em Epistemology of Fundamental Physics}, Cambridge University Press (forthcoming).
\end{center}


\begin{abstract}\noindent
Information theory is increasingly invoked by physicists concerned with fundamental physics, including black hole physics. But to what extent is the application of information theory in those contexts legitimate? Using the case of black hole thermodynamics and Bekenstein's celebrated argument for the entropy of black holes, I will argue that information-theoretic notions are problematic in the present case. Bekenstein's original argument, as suggestive as it may appear, thus fails. This example is particularly pertinent to the theme of the present collection because the Bekenstein-Hawking formula for black hole entropy is widely accepted as `empirical data' in notoriously empirically deprived quantum gravity, even though the laws of black hole thermodynamics have so far evaded empirical confirmation. 
\end{abstract}

\section{Introduction}\label{sec:intro}

Physicists working in quantum gravity diverge rather radically over the physical principles that they take as their starting point in articulating a quantum theory of gravity, over which direction to take from these points of departure, over what we should reasonably take as the goals of the enterprise and the criteria of success, and sometimes even over the legitimacy of different methods of evaluating and confirmating the resulting theories. Yet, there is something that most of them agree upon: that black holes are thermodynamical objects, have entropy and radiate, and thus that the Bekenstein-Hawking formula for the entropy of a black hole must be recovered, more or less directly, from the microphysics of the fundamental degrees of freedom postulated and described by their theories.\footnote{For instance, string theorists hail the result by \citet{strvaf96} of recovering the Bekenstein-Hawking formula from the microphysics of a class of five-dimensional extremal (and hence unrealistic) black holes, advocates of loop quantum gravity celebrate its kinematic derivation (up to a constant) in \citet{rov96}, and proponents of causal set theory applaud the black hole model constructed in \citet{dousor03} in which the black hole's entropy can be seen as proportional to the area of the horizon. All of these results are partial and amount to proofs of concepts, rather than to independently motivated and completely general results. As far as I am aware, Hawking radiation has not been derived in any of the approaches to full quantum gravity. Recently, however, it has been pointed out to me that the temperature of near-extremal holes can be calculated in string theory, as can the absorption and emission rates for stringy quanta, both in agreement with Hawking's calculation.} Thus, regardless of which programme in quantum gravity they adhere to, physicists of all stripes take it as established that black holes are thermodynamic objects with temperature and entropy. Thus, they {\em trust} the Bekenstein-Hawking formula for the latter as a test which a theory of quantum gravity must survive. 

Why do physicists accept the Bekenstein-Hawking result as a litmus test for quantum gravity? The confluence of a number of distinct aspects of known black hole physics all pointing towards thermodynamic behaviour is certainly remarkable. Hawking's calculation in particular uses rather common semi-classical methods of quantum field theory on curved spacetimes. To many it seems hard to imagine that those methods should not be valid at least in some relatively placid arena of mildly semi-classical physics sufficient to derive Hawking radiation. The robustness of these thermodynamic properties of black holes under various theoretical considerations thus inspires a great deal of confidence in them. 

Yet none of these properties has ever been empirically confirmed. So given this lack of direct observations of Hawking radiation or of any other empirical signatures of the thermodynamics of black holes, the universality and the confidence of this consensus is prima facie rather curious. The aim of this article is to investigate the foundation of this seeming, and curiously universal, agreement; in particular, it analyzes Jacob Bekenstein's original argument that led him to propose his celebrated formula, as its recovery constitutes an important assumption of much research into quantum gravity. The unexamined assumption is not worth believing, after all. 

Two caveats before we start. First, if non-empirical theory confirmation is legitimate, as has been suggested e.g.\ by Richard \citet{daw13}, then the fact that thermodynamic behaviour of black holes has not been observed to date may not be a worry as long as we have convincing non-empirical reasons for believing the Bekenstein-Hawking formula. In revisiting Bekenstein's original argument, I propose to evaluate some of these reasons. Second, I appreciate that the case for taking the Bekenstein-Hawking formula seriously would be significantly strengthened by the observation of Hawking radiation in models of analogue gravity.\footnote{As argued, e.g., by \citet{dareal17}. \citet{ste16} claims to have observed Hawking radiation in analogue black holes in a Bose-Einstein condensate, but to my knowledge the result has not been replicated.} In fact, until Stephen \citet{haw75} offered a persuasive semi-classical argument that black holes radiate, and so exhibit thermodynamic behaviour like a body with a temperature, most physicists were not moved by Bekenstein's earlier case for black hole entropy. I will briefly return to this point in the conclusions, but leave the detailed analysis for another occasion. 

The next section covers some of Bekenstein's original motivations and discusses the area theorem in general relativity (GR), which serves as the vantage point for Bekenstein. Section \ref{sec:argument} offers a reconstruction of Bekenstein's main argument. Section \ref{sec:salience} evaluates Bekenstein's attempt to endow the central analogy with physical salience and challenges his invocation of information theory to this end. Section \ref{sec:conc} concludes.

\section{The generalized second law and the area theorem}\label{sec:second}

\citet{bek72} is motivated by the recognition that black hole physics potentially violates the Second Law of thermodynamics, which states that the entropy of an isolated physical system never decreases, but always remains constant or increases.\footnote{For an authoritative and philosophical introduction to black hole thermodynamics, see \citet[\S5]{cur17a}.} He considers the case of a package of entropy lowered into a black hole from the asymptotic region and asserts that once this has settled into equilibrium,
\begin{quote}
...there is no way for the [exterior] observer to determine its interior entropy. Therefore, he cannot exclude the possibility that the total entropy of the universe may have decreased in the process... The introduction of a black-hole entropy is necessitated by [this] process. (737f)
\end{quote}
Thus, Bekenstein contends that once the entropy package has vanished behind the event horizon of the black hole and thus becomes in principle unobservable, it may well be, for all we know, that this entropy has dissipated and thus that the total entropy in the universe has decreased, {\em pace} the Second Law. One might choose to resist his argument right here: in a classical account at least, any observer in the asymptotic region will in fact never see the entropy package actually disappear behind the event horizon and so for all they can tell, it will always remain there, and thus will in no way threaten the Second Law.\footnote{I wish to thank Niels Linnemann for pointing this out to me.}

Suppose, however, that the asymptotic observer has reason to believe that the entropy package has been swallowed by the black hole's event horizon; perhaps the observer knows general relativity and calculates that the package has in fact passed the event horizon, even though this is not what they see. So they could have reason to believe that the Second Law is violated. This problem would obviously be avoided if the black hole itself possessed entropy, and moreover possessed an entropy that would increase by at least the amount of entropy that disappears behind the event horizon. The ascription of entropy to a black hole\footnote{Entropy is standardly predicated only of systems in equilibrium. The standard black hole solutions in classical general relativity are stationary and in this sense in equilibrium; however, not every physical black hole is thought to be in equilibrium, and so the ascription of entropy to {\em those} black holes would require non-standard thermodynamics.} thus permits a generalization of the traditional Second Law, as Bekenstein asserts:
\begin{quote}
{\em The common entropy in the black-hole exterior plus the black-hole entropy never decreases.} (\citeyear{bek73}, 2339, emphasis in original)
\end{quote}
At this point in the argument, the only justification for introducing black hole entropy rests on saving (a generalized version of) the Second Law. Although this is not the main focus of the present essay, it is important to realize that this amounts to a rather significant extension of the validity of the Second Law from the firm ground of terrestrial physics to the precarious and speculative realm of black holes. For all the evidence we have, black holes may well exhibit unusual and unexpected thermodynamic behaviour---or none at all. It is certainly not automatic that the Generalized Second Law holds. In fact, not only are many proofs of the Generalized Second Law lacking \citep{wal09}, but Aron \citet{wal13} has shown that a singularity theorem results from it. Just as energy conditions constitute essential premises for the standard singularity theorems of classical GR, the Generalized Second Law seems to play this role for a new version of a singularity theorem, which no longer depends on energy conditions, and thus may hold well into the quantum realm. Thus, that something like the Generalized Second Law obtains is far from trivial, particularly if we think that the presence of singularities points to a failure of our theories. 

Once we accept it, however, and with it the need for a black hole entropy, we are faced with the question of what the black hole entropy is and how it relates to other physical properties of black holes. \citet{bek72} finds the answer in a theorem that Hawking proved just a year earlier \citep{haw71}: the area theorem. In the formulation of Robert \citet[138]{wal94}, the area theorem is the following proposition:
\begin{thm}[Area theorem (Hawking 1971)]\label{thm:area}
For a predictable black hole satisfying $R_{ab} k^a k^b \geq 0$ for all null $k^a$, the surface area of the future event horizon, $h^+$, never decreases with time.
\end{thm}
Here, $R_{ab}$ is the Ricci tensor, $k^a$ a null vector tangent to the geodesic generators of the event horizon, and $h^+$ the future event horizon of the black hole, i.e., the boundary of all events in the spacetime that can send light signals to infinity. The surface area of the horizon $h^+$ is understood as the area of the intersection of $h^+$ and some spacelike (or possibly null) hypersurface $\Sigma$ suitable to evaluate the entropy. For a proof of the theorem, see \citet[138f]{wal94}.

The area theorem seems to give us an obvious candidate for a physical quantity to act in lieu of standard entropy: the never-decreasing surface area of the black hole's event horizon. Bekenstein argues for the explicit analogy of the area theorem and the Second Law, and the roles played in them, respectively, by the area surface of the event horizon and the black hole's entropy. Although the propositions remain merely analogous, the intended connection at the level of the physical magnitudes is one of identity: the area of the event horizon really {\em is} the black hole's entropy. Or so goes the thought. It is my main purpose in this paper to analyze Bekenstein's argument for this connection. While Bekenstein's argument may not be the most decisive of arguments in favour of attributing an entropy to black holes, it gave birth to an industry and thus deserves scrutiny. 

Before we analyze the premises of the theorem, it ought to be noted that the analogy breaks down right out of the gate, at least in some respects. At least if we accept its nomic reduction to statistical mechanics, then the Second Law is not a strict law, but is only statistically valid: it suffers from occasional, though very rare, violations. Not so for the area theorem, which is a strict theorem in the context of GR: the area of the event horizon {\em never} decreases. This means that the time asymmetry grounded in the area theorem is also strict, whereas the time asymmetry based on the Second Law is not strict---again, at least if we accept that it is nomically reducible to statistical mechanics. Now while this is clearly a disanalogy, it may not be fatal to Bekenstein's endeavour: it may be that the physics of black holes is different from that of ordinary thermodynamic objects in that there just is no possibility of decreasing entropy; or it may be that the area law, just like the Second Law, will have to be replaced by a proposition in a more fundamental (quantum) theory of gravity that will permit rare exceptions to the increase of area. This second possibility is, of course, just what many physicists working in quantum gravity expect. In fact, Hawking radiation and the evaporation of black holes, if borne out, hands us a mechanism counteracting the exceptionless increase of entropy suggested by the area theorem. Hawking radiation thus renders the area theorem otiose, but this should not come as a surprise given that the area theorem is a theorem in classical GR and Hawking radiation a result in semi-classical gravity. This suggests that we should not expect the area theorem to hold strictly once we move beyond the domain of GR. 

Either way, what really matters is the Generalized Second Law, not the area theorem, which is, to repeat, just a theorem in classical GR. Whether the Generalized Second Law holds, approximately or strictly or not at all, once quantum effects are included is a rather involved question.\footnote{Cf.\ \citet[\S 4.2]{wal01} for a discussion. Note that the loss of black hole entropy in the form of a shrinking area is thought to be compensated by the ordinary entropy of the escaping Hawking radiation.} So insisting that Bekenstein's argument breaks down already at this early stage may be taking the area theorem too seriously as a statement about the fundamental behaviour of black holes. 

A further note of interest before we proceed. Insofar as we pick a highly improbable initial state, there is a sense in which the time asymmetry coming out of the Second Law is put in by hand. But the asymmetry of time grounded in the area theorem is no better off: it is equally stipulated when we demand that the horizon is a {\em future}, rather than a past, horizon; in other words, when we postulate that the singularity forms a {\em black}, rather than a {\em white}, hole.\footnote{For the distinction between black and white holes, see \citet[155]{walgr}.} In fact, the situation may be worse in the case of the area theorem: in the case of the Second Law, the existence of a dynamical process that {\em drives} the early universe toward such a state would overcome this unsatisfactory sleight of hand and give us a physical mechanism that renders stipulating a highly special initial state obsolete. 

The area theorem, like any other theorem, rests on premises. The two central assumptions are that of `predictability' and a positivity condition on the Ricci tensor. Following \citet[137]{wal94}, predictability is defined as follows:
\begin{defi}[Predictable black hole]
A black hole in a spacetime $\langle\mathcal{M}, g_{ab}\rangle$ is called {\em predictable} just in case it satisfies the following conditions, where $\Sigma$ is an asymptotically flat time slice, $I^\pm(X)$ is the chronological future/past of the set $X$ of events in $\mathcal{M}$, $\mathscr{I}^+$ the future null boundary of $\mathcal{M}$, and $D(X)$ the domain of dependence of the set $X$:
\begin{enumerate}
\item[(i)] $[I^+(\Sigma) \cap I^-(\mathscr{I}^+)] \subset D(\Sigma)$,
\item[(ii)] $h^+ \subset D(\Sigma)$.
\end{enumerate}
\end{defi}
Equivalently, an asymptotically flat spacetime $\langle \mathcal{M}, g_{ab}\rangle$ comprises a predictable black hole if the region $\mathcal{O}\subset \mathcal{M}$ to the future of $\Sigma$ containing the region exterior to the black hole as well as the event horizon $h^+$ is globally hyperbolic.  This explains the choice of terminology: in a spacetime containing a predictable black hole, all events outside the black hole and to the future of time slice $\Sigma$ are `visible' from (future null) infinity, and, generically, the physics in the region exterior to the black hole is `predictable' in the sense that typical dynamical equations for sufficiently well-behaved matter such as the Klein-Gordon equation have a well-posed initial value problem \citep[Theorem 4.1.2]{wal94}. This means that these spacetimes satisfy a version of the Cosmic Censorship Hypothesis (ibid., 135). Whether or not the relevant version of the Cosmic Censorship Hypothesis (or, for that matter, any version of it) is true, is assumed, but not known, and a matter of ongoing research \citep{pen99}. Without dwelling on the matter though, it should be noted that this may well be a non-trivial limitation of the scope of the area theorem. 

Matters are worse with respect to the second substantive premise of the area theorem. The positivity requirement on the Ricci tensor, $R_{ab} k^a k^b \geq 0$, is related to the so-called `null energy condition':
\begin{defi}[Null energy condition]
$T_{ab} k^a k^b \geq 0$, with $k^a$ any future-pointing null vector field, where $T_{ab}$ is the stress-energy tensor.
\end{defi}
If the Einstein field equation holds, then the null energy condition obtains just in case $R_{ab} k^a k^b \geq 0$ obtains. So in GR, the null energy condition implies and is implied by the positivity condition on $R_{ab}$. Since there exists a long list of possible violations of the null energy condition \citep[\S 3.2]{cur17}, the second premise of the theorem constitutes a more severe limitation. The theorem simply does not speak to those spacetimes which either contain a non-predictable black hole or where the null energy condition is violated; in those cases, even if the entropy of a possibly present black hole is proportional to the surface area of its event horizon, we have no guarantee that the area and hence the black hole entropy will never decrease. 

One might object to my reservations here that the situations in which we are faced with a non-predictable black hole or a violation of the null energy condition, while formally possible, are not in fact physically possible. I accept that we may ultimately come to this conclusion. However, we should be wary to rule out those cases based on not much more than our intuitions concerning what is physically possible. GR is a notoriously permissive theory; but also a notorious case for challenging our physical intuitions. For example, the existence of singularities in some of its models has undergone a transformation from being so unpalatable as to be ruled out {\em by fiat} to a necessary part of celebrated predictions of the theory. The challenge in GR is to distinguish cases when it is overly permissive in that it allows unphysical solutions from those situations in which it holds genuinely novel, perhaps unexpected, and potentially important lessons which we should heed. In this spirit, it is important to keep the limitations of the area theorem in mind as we proceed.

\section{The structure of Bekenstein's argument}\label{sec:argument}

Bekenstein's argument to the conclusion that black holes have entropy, and that this entropy is proportional to the area of the black hole's event horizon,\footnote{\citet{bek72} only argues for the proportionality of entropy and area. Although he tries to fix the proportionality factor in \citet{bek73}, the generally accepted value of 1/4 is first proposed in \citet{haw75}.} essentially contains two parts. First, it establishes a formal similarity between thermodynamic entropy and black-hole entropy. As this formal analogy is, in itself, rather weak, a second step establishing the physical salience of the formal similarity is required. 

In \citet{bek72}, the formal similarity is noted, and it is argued that if we use it to endow black holes with entropies proportional to the surface area of their event horizon (in Planck units), then the appropriately reformulated Second Law will still be valid for processes involving black holes that would otherwise apparently violate it. Thus, the first step in the argument is essentially present in the 1972 paper. Although it also contains inklings of the information-theoretic justification of the second step, the full defence of that is only given in \citet{bek73}. Let us analyze the two steps in turn, starting with the first. 

As \citet{bek72} notes, it appears as if black holes permit interactions with their environment such as lowering entropy packages into them which apparently result in a decrease of thermodynamic entropy, thus violating the Second Law. Thermodynamic entropy is an additive state variable and gives a measure for the irreversibility of the evolution of an isolated thermodynamic system. The entropy of such a system remains constant just in case the isolated system evolves reversibly; it increases if and only if it evolves irreversibly. This gives us the entropic form of the Second Law, according to which the entropy of an isolated system cannot decrease.\footnote{For a useful introduction to entropy, the reader may consult \citet{lem13}.} Quantitatively, a differential increment of thermodynamic entropy $S_{TD}$ is given by a form of the Second Law,
\begin{equation}
dS_{TD} = \frac{\delta Q}{T},
\end{equation}
where $\delta Q$ is an indefinitely small quantity of heat that is transferred to or from the system, and $T$ is its absolute temperature. In order to block the possibility that heat can be converted to work without attendant compensating changes in the environment of the system, Bekenstein generalizes the Second Law to include a term for the entropy of black holes, as outlined in \S\ref{sec:second}. He then identifies the area of the horizon of a black hole as the only physical property of a black hole that, just like thermodynamic entropy, exhibits a tendency to increase, as shown by Theorem \ref{thm:area}:
\begin{quote}
The area of a black hole appears to be the only one of its properties having this entropylike behavior which is so essential if the second law as we have stated it is to hold when entropy goes down a black hole. \citep[738]{bek72}
\end{quote}
Thus, the sole basis on which Bekenstein introduces the claim that the area of its event horizon is (proportional to) a black hole's entropy is that it is the only property that is not obviously altogether {\em unlike} entropy. A weak similarity indeed. 

Given the parallel tendency to increase, Bekenstein proposes that the entropy of a black hole $S_{BH}$ be simply given as proportional to the area $A$ of its horizon:
\begin{equation}\label{eq:bhentropy1}
S_{BH} = \eta \frac{kA}{\ell_P^2},
\end{equation}
where the introduction of Boltzmann's constant $k$ and the Planck length $\ell_P$ is ``necessitated by dimensional considerations'' (\citeyear[738]{bek72}).\footnote{More precisely, they are necessitated by dimensional considerations together with an argument that these are ``the only truly universal constant[s]'' (ibid.) of appropriate dimension.} $\eta$ is a constant of order unity, and part of Bekenstein's ambition in his longer article of 1973 was to fix that proportionality factor; he ends up suggesting $\eta = \ln 2/8\pi$, or something ``very close to this, probably within a factor of two'' (\citeyear[2338]{bek73}). This number is, however, an order of magnitude smaller than Hawking's generally accepted value of 1/4, which then yields what has since become known as the `Bekenstein-Hawking formula' for the entropy of a black hole:
\begin{equation}\label{eq:bhentropy2}
S_{BH} = \frac{kA}{4\ell_P^2}.
\end{equation} 

At this point, then, there is a tenuous formal similarity between entropy and horizon area underwritten only by the fact that they both tend to increase; from this formal similarity, equation (\ref{eq:bhentropy1}) is divined as a simple choice with the right dimensions if the analogy were physical. There are plenty of examples of formal analogies that we should not reasonably think of as amounting to more than just that: merely formal analogies. For instance, the mathematics of the Lotka-Volterra equations can be used to model the `predator-prey dynamics' of ecosystems of populations consisting of two species, just as they can describe the dynamics of two macroeconomic variables, the workers' share of the national income and the employment rate.\footnote{\citet[77f]{wei13} considers Goodwin's appropriation of the Lotka-Volterra equations for this macro\-economic model a strong example of ``construal change'' (ibid.) of a model. Note that Goodwin took the analogy to be more than merely formal; but in this, he was arguably motivated by his Marxism, illustrating just how the interpretation of the salience of an analogy may depend on one's ideological stance.} One of the most dramatic examples of a formal analogy relates the resonances in the scattering spectrum of neutrons off of atomic nuclei and the frequencies of large-scale oscillations in the distribution of prime numbers through the integers described by the zeroes of Riemann's zeta function.\footnote{Cf.\ \citet{tao12}; Tao gives other examples of such `universal' behavior.} That these analogies are purely formal should be clear; scattering neutrons and distributing prime numbers are hardly subject to the same laws of nature, and ecology and macroeconomics are arguably governed by different forces. In the light of this, why should one take the formally even weaker analogy between horizon area and thermodynamic entropy seriously?

\begin{figure}[t]
\centering
\epsfig{figure=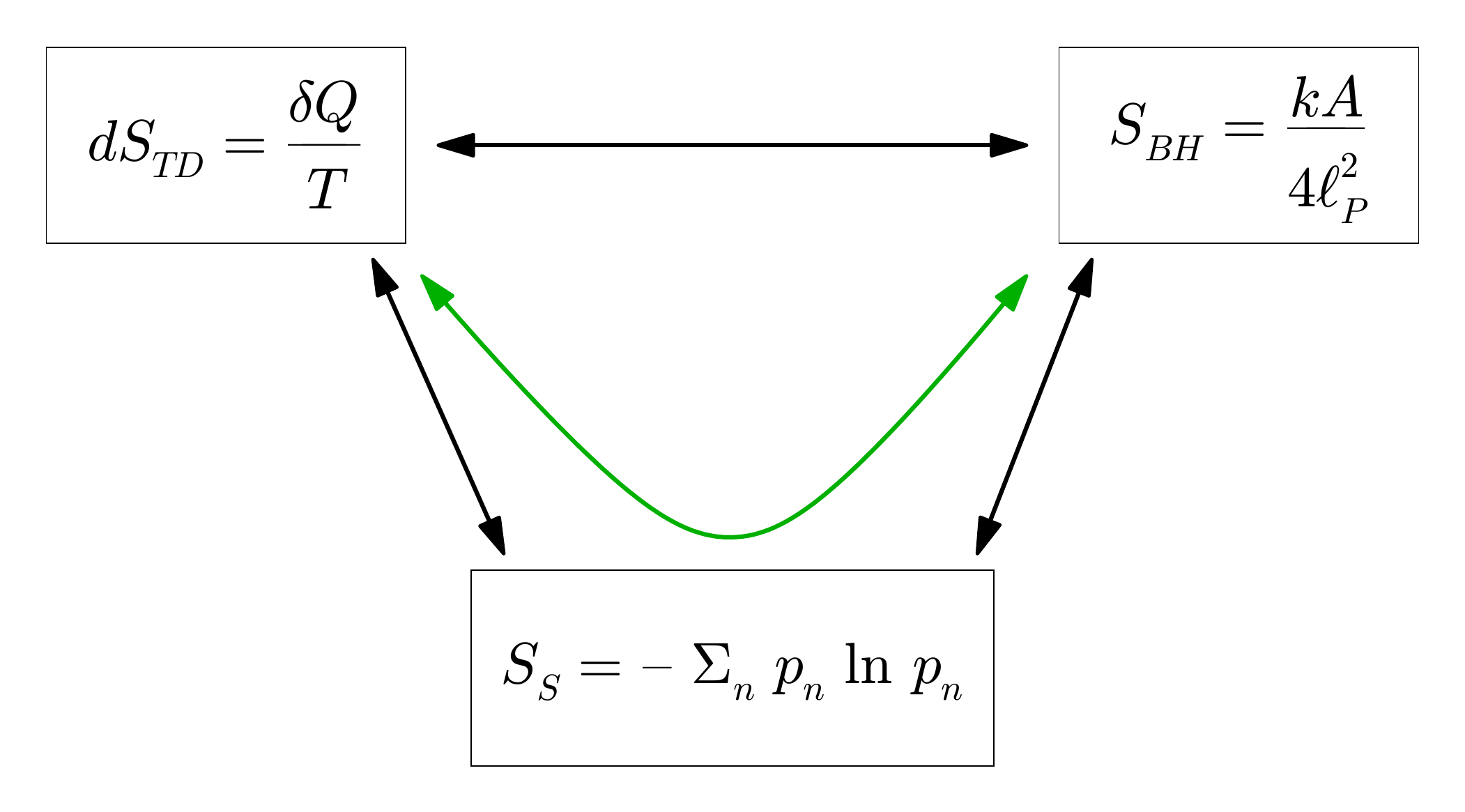,width=0.75\linewidth}
\caption{\label{fig:argstructure} The green path through the information-theoretic concept of entropy (bottom) shows the inferential structure of Bekenstein's claim of physical salience of the analogy.}
\end{figure}

I read \citet{bek73} as an attempt to make the case for a more robust analogy, a {\em physical} analogy. His argument to this effect---itself the second part of the overall argument as described at the beginning of this section---is best analyzed as consisting of two parts. First, the thermodynamic entropy of isolated systems is identified with Claude Shannon's information-theoretic entropy. Second, the black-hole area gets identified with the information-theoretic concept of entropy. Both steps combined then amount to an identification of the area of the black hole's event horizon with thermodynamic entropy. Since both steps do not just formally identify, but in fact carry physical salience across the identifications, the area of the horizon {\em really is} an entropy akin to the usual thermodynamic entropy. The inferential path through the information-theoretic concept of entropy is thus what establishes the physical salience (cf.\ Figure \ref{fig:argstructure}). Of course, this identification then underwrites the generalization of the Second Law: as black-hole entropy is of the same physical kind as ordinary thermodynamic entropy, they can unproblematically be summated to deliver the total entropy of the universe.

In sum, Bekenstein's argument can be broken down as follows:
\renewcommand{\labelenumi}{(\roman{enumi})}
\begin{enumerate}
\item The universe as a whole has an entropy.
\item The Second Law is a universal law, i.e., applies to the universe as a whole; in particular (a generalization of) it still obtains in the presence of black holes. 
\item The Second Law can only hold in the presence of black holes if the black holes present have an entropy contributing to the total entropy of the universe.
\item The area of the event horizon of a black hole is the only property of a black hole which is formally similar to ordinary thermodynamic entropy in that they both tend to increase over time.
\item The horizon area of a black hole is essentially the same kind of physical quantity as information-theoretic entropy.
\item Information-theoretic entropy is essentially the same kind of physical quantity as ordinary thermodynamic entropy.
\item Therefore, the horizon area of a black hole and ordinary thermodynamic entropy are essentially the same kind of physical quantity (which explains their formal similarity).
\item Therefore, black holes have an entropy contributing to the total entropy of the universe.
\item Therefore, (a generalization of) the Second Law still obtains in the presence of black holes.
\end{enumerate}
I take it that steps (vii) and (viii), assuming the transitivity of `being essentially the same kind of physical quantity', are harmless. Steps (iii) and (ix) are perhaps not entirely automatic, but shall be granted for present purposes. One might argue that for claim (iii) to hold, an information-theoretic conception of entropy (and, moreover, an epistemic reading of that conception) is required for Bekenstein's motivating argument that otherwise, the Second Law may be violated, to have any force at all. Alternatively, one might rephrase (iii) as follows in order to be clear what that argument can establish at most:
\begin{enumerate}
\item[(iii$'$)] {\em We can only be certain that} the Second Law continues to hold in the presence of black holes if the black holes present have an entropy contributing to the total entropy of the universe.
\end{enumerate}
While this complaint is perfectly justified, the point of my reconstruction is to isolate the move to an (epistemic reading of an) information-theoretic conception of entropy in a later step in the argument---in the hope to make the problematic nature of this step more evident. So let me accept (iii) as it stands for now. 

Similarly to (iii) and (ix), (i) is a substantive assumption that may turn out to be false, or misguided; but we shall grant it for now, too. We have already seen in \S\ref{sec:second} that (ii) is not unproblematic as black-hole physics may well violate the Second Law. In the same section, we have also found that (iv) is not strictly true. Putting concerns regarding (ii) and (iv) to the side though, we are left with the most interesting claims of Bekenstein's argument: Steps (v) and (vi), which are designed to secure the physical salience of the analogy. Before we address them in the next section, it should be noted that for both (v) and (vi), they could be interpreted in different, inequivalent, ways. Both statements are of the form `Thing $A$ is essentially the same kind as thing $B$'. While Bekenstein and others often seem to think of these analogies as identities, strictly weaker statements of the form `Thing $A$ is essentially one of the things that make up the kind of things $B$', leaving open the possibility that the second placeholder designates a more general and encompassing category, would suffice for the argument to go through. I take it that these weaker statements would leave the transitivity necessary to infer to (vii) intact. One thing that is clear, though, is that the inverse direction would not work: it would not suffice to establish that thing $B$ is a proper subset of the things that make up kind $A$. 

\section{The quest for physical salience}\label{sec:salience}

In \S2 of his often cited paper of 1973, Bekenstein offers three formal analogies between black-hole physics and thermodynamics. First, as already noted, both the surface area of black-hole horizons and thermodynamic entropy tend to increase over time. Second, in thermodynamics, entropy can be considered a measure of the degradation of the energy of a system, in the sense that degraded energy cannot be transformed into work: if a thermodynamic system reaches maximum entropy it can no longer be used to extract work. Surprisingly, if one considers interactions with a black hole by so-called `Penrose processes' on black holes with angular momentum, one can lower the mass (and the angular momentum) of the black hole, thereby extracting energy, up to a certain maximum.\footnote{Penrose processes were originally articulated in \citet{pen69}; cf.\ also \citet[\S 12.4]{walgr}.} Energy can be extracted from such black holes until their angular momentum is halted and the extractable energy thus is depleted. The `irreducible mass' $M_{irr}$ of a black hole represents the  non-extractable mass-energy of the black hole. For a Kerr black hole, its square turns out to be proportional to the horizon area \citep[eq.\ 12.4.12]{walgr}:
\begin{equation}\label{eq:irrmass}
A = 16\pi M_{irr}^2.
\end{equation}
Thus, the irreducible mass of a black hole acts similarly to the maximum entropy a thermodynamic system can obtain. Throughout a Penrose process where the mass of the black hole is held constant, its angular momentum decreases and its irreducible mass increases. Thus, its area as given by equation (\ref{eq:irrmass}) increases, in concordance with the Area Theorem---until the Kerr black hole is `spun down' to a Schwarzschild black hole of zero angular momentum, at which point no more energy can be extracted from it.\footnote{I thank Erik Curiel for clarifications on this point.}

The third formal analogy noted by Bekenstein arises from the possibility of finding a black-hole analogue of the so-called `fundamental constraint' imposed by the first two laws in the thermodynamics of simple fluids:
\begin{equation}
dE = T dS - P dV,
\end{equation}
where $E$ is the internal energy, $T$ the temperature, $S$ the entropy, $P$ the scalar pressure, and $V$ the volume of the fluid \citep[\S1.9]{lem13}. The analogy is tenuous, however, at least for black holes as described by general relativity, as it requires a black-hole analogue of temperature $T$. Although such an analogue can formally be defined in classical general relativity as a function of the black hole's three parameters---its total mass $M$, its total angular momentum per mass $a$, and its total electric charge $Q$---, the interpretation of this magnitude as a physical temperature is unavailable as classical black holes cannot radiate and as it thus makes little sense to attribute temperatures to them. Of course, this changes once quantum effects are taken into account. We will return to the question of the quantum nature of black hole thermodynamics in the concluding section.

Let us start with proposition (vi) in the reconstruction of Bekenstein's argument in the previous section, the association of information-theoretic entropy with ordinary thermodynamic entropy. First, assuming a nomic reducibility of thermodynamics to statistical physics, the macroscopic entropy for an ideal gas gets replaced with the microscopic {\em Boltzmann entropy} defined as 
\begin{equation}\label{eq:boltz}
S_B = k \log W,
\end{equation}
where $k$ is, as above, the Boltzmann constant and $W$ is the `size' of the given macrostate in the  phase space of admissible microstates or its `multiplicity' (or ``Permutabilit\"at'' in Boltzmann's own words), i.e., the number of microstates corresponding to the macrostate at stake. Thus, the Boltzmann entropy in (\ref{eq:boltz}) presupposes an atomic hypothesis according to which the isolated macroscopic thermodynamic system really consists of microscopic parts. It furthermore presupposes that the microstates of the system, i.e., the fundamental states of its parts, are all equally probable. The microstates give rise to the system's `macrostates' defined in terms of observable, macroscopic physical properties of the system. Importantly, the macrostates are multiply realized by the microstates. 

It should be noted that already this first identification of thermodynamic entropy with a statistical notion is substantive in that the two entropies are clearly conceptually, and sometimes---when out of equilibrium---numerically, distinct.\footnote{Cf.\ e.g.\ \citet{cal99}.} Consequently, there have been efforts, e.g.\ by \citet{goumay01} and \citet{prutim17}, to establish the thermodynamic behaviour of black holes and to derive the Bekenstein-Hawking formula based on thermodynamic concepts and assumptions alone and thus cleanse the arguments from their usual detour through statistical notions. While such efforts establish that thermodynamic behaviour can consistently be attributed to black holes, they do not show this from independent assumptions about black hole physics: they {\em presuppose} that the laws of thermodynamics apply to black holes---and they assume that they emit Hawking radiation. Let's return to Bekenstein's original argument. 

The fundamental postulate of statistical mechanics, a presupposition of equation (\ref{eq:boltz}), ascertains that the microstates of an isolated thermodynamic system are all equally probable. For systems interacting with their environment, the equiprobability of their microstates cannot be assumed, and a generalization of (\ref{eq:boltz}) to thermodynamic systems with microscopic degrees of freedom assuming microstates of possibly unequal probability is required. The {\em Gibbs entropy} delivers just that:
\begin{equation}\label{eq:gibbs}
S_G = - k\; \Sigma_i\, p_i \log p_i,
\end{equation}
where $p_i$ is the probability that the $i$-th microstate obtains.\footnote{Strictly speaking, the Gibbs entropy is defined on the basis of an {\em ensemble} of a large number---eventually an infinity---of independent copies of the given system. We will ignore this here. We will also ignore the more general case of continuously many microstates.} Comparing the Gibbs entropy of a thermodynamic system with  Shannon's dimensionless ``measure of uncertainty'' which quantifies the uncertainty lost by a message,
\begin{equation}\label{eq:shannon}
S_S = - \Sigma_n p_n \log_2 p_n,
\end{equation}
we find another formal analogy: the right-hand side of equation (\ref{eq:gibbs}) is proportional to that of equation (\ref{eq:shannon}).\footnote{Assuming, that is, binary states; this assumption is standard in information theory, where information is measured in bits, but also common in statistical mechanics.} 

To repeat, in Shannon's information theory, $S_S$ quantifies uncertainty reduced by the transmission of a message, or the attenuation in phone-line signals. Why should anyone think that it is essentially the same kind of physical quantity as thermodynamic entropy, originally introduced as a measure of the irreversibility of the dynamical evolution of isolated systems such as boxed-off ideal gases? Shannon himself, in a now famous recollection, testifies to the shaky origins of the association of his new formula with thermodynamic entropy:
\begin{quote}
My greatest concern was what to call it. I thought of calling it `information', but the word was overly used, so I decided to call it `uncertainty'. When I discussed it with John von Neumann, he had a better idea. Von Neumann told me, `You should call it entropy, for two reasons. In the first place your uncertainty function has been used in statistical mechanics under that name, so it already has a name. In the second place, and more important, nobody knows what entropy really is, so in a debate you will always have the advantage.' \citep{tribus71}
\end{quote}
Thus, the first goal of a successful argument establishing the essential physical kinship of a black hole's horizon area and thermodynamic entropy must be to physically substantiate the association apparently so frivolously made by von Neumann. Attempts to do so predate Bekenstein's work by more than fifteen years and so ensue shortly after the publication of Shannon's field-defining classic of 1948. Based on Shannon's central insight that the new information received by the epistemic agent corresponds to the decrease of the agent's uncertainty about a system's internal state, L\'eon \citet{bri56} identifies information with negative entropy. In the late 1950s and 1960s, Brillouin, Edwin T Jaynes, Rolf Landauer, and others attempt to solidify the association between information theory, statistical physics, and thermodynamics. For instance, \citet{jay57a,jay57b} first derives statistical mechanics from information-theoretic assumptions and \citet{lan61} asserts that the erasure of information requires a minimum amount of energy of $kT \ln 2$ per bit of information erased, where $T$ is the temperature of the circuit.\footnote{Cf.\ \citet{mar09} for a critical introduction to the relationship between information processing or computation and thermodynamics.} 

The literature on this topic is vast and certainly not conclusive. I shall not attempt to do it justice here and only give a few pointers. For instance, there have been claims of an experimental confirmation of `Landauer's principle' \citep{bereal12}---refuted by John \citet[\S3.7]{nor13}---, advocacy of the principle \citep{ben03} as well as criticism of it \citep{nor11}. Some physicists have gone as far as claiming that the entire material universe is nothing but information.\footnote{\citet{ved10}, for example, is a popular book advocating this. One of the founding fathers of the physics-is-information school was of course John Archibald Wheeler---Bekenstein's PhD advisor! \citet{whe90} is arguably his most pronounced articulation of the idea that physical objects depend in their very existence on information, as captured in the slogan `It from bit'.} In the opposite direction, Norton warns that the focus on information theory has produced an identifiable harm in the context of Maxwell's demon: information-based arguments against the demon have not only been unsuccessful, but have furthermore precluded us from recognizing the straightforward violation of (the quantum analogue of) Liouville's theorem as a definitive exorcism of the demon.\footnote{Cf.\ \citet[\S4]{nor13} for the classical argument and \citet{nor18} for the extension to the quantum case.}

One problem of the information-theoretic approach, also noted by Owen \citet{mar09}, is that radically different concepts and standards of proof are operative in this literature, collectively obscuring the synthesizing lesson to be drawn from it. Although it remains popular among physicists, there are excellent reasons to resist the identification of thermodynamic entropy, which is grounded in objective facts about the microphysics of physical systems, with information-theoretic entropy, which concerns subjective facts about the epistemic state of agents, as John Dougherty and Craig Callender \citeyearpar[\S5]{doucal17} insist. In fact, the two entropies can have distinct numerical values when applied to the same system at the same time. As an example, they offer an isolated ideal gas that has whatever thermodynamic entropy it has regardless of whether or not we know its precise microstate:
\begin{quote}
If a Laplacian demon told you the exact microstate of a gas, that would affect the value of the Shannon entropy (driving it to zero) whereas it wouldn't affect the value of the Gibbs entropy. (20f)
\end{quote}
Thus, they conclude, it would be a big mistake to identify the two concepts. 

I will not dwell on the profound issue of the relationship between physics and information theory here. However, it must be noted that in order for Bekenstein's argument to succeed, it is necessary that all information-theoretic entropy essentially be thermodynamic entropy, as asserted in (vi). As noted at the end of the previous section, while an identity of the two concepts would certainly establish (vi), it would dialectically suffice if information-theoretic entropy would turn out to be just one sort, or one manifestation, of thermodynamic entropy among others. It would not suffice, however, if it turned out, as is generally assumed, that Shannon entropy is more general than thermodynamic entropy in the sense of also assigning (Shannon) entropy to systems without thermodynamic entropy. This would not be sufficient precisely because it could then turn out that while black-hole entropy is indeed information-theoretic entropy, it is of the non-thermodynamic kind. If information-theoretic entropy is a more general concept than thermodynamic entropy, then we would not only have to establish that black-hole entropy is information-theoretic, but also that it is furthermore of the thermodynamic kind. This means that general proofs \`a la Jaynes that statistical mechanics can be asymmetrically reduced to information theory will not do all the work asked of proposition (vi) in my reconstruction of Bekenstein's case. 

It should be noted that Bekenstein explicitly endorses what could be called the `Jaynesian identification':
\begin{quote}
The entropy of a system measures one's uncertainty or lack of information about the actual internal configuration of the system... The second law of thermodynamics is easily understood in the context of information theory... (\citeyear[2335]{bek73})
\end{quote}
Again, to subsume both the second law of thermodynamics and black hole physics into information theory will still require an argument as to why thermodynamic entropy and horizon area are relevantly similar---other than both being merely (and perhaps in relevantly distinct ways) information-theoretic. 

The upshot of all this is that it would be inadmissibly facile to think that proposition (vi) is firmly established. It isn't, as there remain open questions regarding the relationship between information theory and thermodynamics. But let us turn to proposition (v), the claim that the horizon area of a black hole is essentially the same kind of thing as information-theoretic entropy. The association between the horizon area and information-theoretic entropy articulated in proposition (v) is Bekenstein's original contribution and a main step in the overall argument. 

Bekenstein sets the tone right away; in the abstract, he writes
\begin{quote}
After a brief review of the elements of information theory, we discuss black-hole physics from the point of view of information theory. We show that it is natural to introduce the concept of black-hole entropy as the measure of information about a black-hole interior which is inaccessible to an external observer... The validity of [the generalized] second law is supported by an argument from information theory... (\citeyear[2333]{bek73})
\end{quote}
In the main body of the text, the theme is amply replayed:
\begin{quote}
In the context of information a black hole is very much like a thermodynamic system... Black holes in equilibrium having the same set of three parameters may still have different `internal configurations'... It is then natural to introduce the concept of black-hole entropy as the measure of the {\em inaccessibility} of information (to the exterior observer) as to which particular internal configuration of the black hole is actually realized in a given case. At the outset it should be clear that the black-hole entropy we are speaking of is {\em not} the thermal entropy inside the black hole. (1973, 2335f, emphases in original) 
\end{quote}
These quotes betray the same misconception: fundamental physics is about the objective structure of our world, not about our beliefs or our information. 

In fact, the problem may be considered to run deeper: information, arguably, is an inadmissible concept in fundamental physics. For there to be information in the first place, there must be a communication system in place, a physical set-up such that the concept of information is applicable. In Shannon's mathematical theory of communication \citep{sha48}, for there to be communication, there must be an {\em information source} of a message, a {\em transmitter} sending a signal, via a potentially {\em noisy channel}, to a {\em receiver}, which receives the signal and decodes it for the {\em destination}. This destination, according to Shannon, is ``the person (or thing) for whom the message is intended'' \citep[381]{sha48}. Even subtracting the intentionality, and abstracting from the personhood of the destination, we are still left with an ineliminable minimum level of complexity required for the signal to be interpreted as the transmission of {\em information}. As Christopher \citet[Ch.\ 2]{tim13} shows, since `information', in Shannon's theory, is ``what is produced by a source that is required to be reproducible at the destination if the transmission is to be a success'' (42), we need to distinguish the abstract type of information from the tokens of its concrete physical instantiations. Hence, `information', as understood in information theory, is an abstractum.\footnote{For a discussion of different intepretative options concerning Shannon entropy and for the defence of an ulimately pluralist stance, cf.\ particularly \citet{lomeal16}. It should be noted that they also insist that Shannon information is not an absolute magnitude, but relative to an entire communication situation (\S9).} In addition to the category mistake then, the complexity of the set-up should not be required of a physical system described by a fundamental physical theory, which is, in general, too impoverished to incarnate all these roles at once at their level. 

Moreover, unless one accepts an ultimately Platonist---indeed Pythagorean---ontology of the material world, fundamental physics ought not be grounded in abstracta like information. The recently popular metaphor of the universe `being' a quantum computer can only remain a claim of {\em physics} if it is interpreted to mean that the universe is a physical system with a nature that is ultimately best described by a quantum theory, perhaps with a quantum dynamics that produces the `steps' of the computing. A physical system is qualitatively distinct from an abstract object or structure. The more systematic assessment of the merits and limitations of an information-theoretic abstraction of fundamental physics shall be left for another day, but it should be clear that if we reject radical Pythagoreanism, i.e., the belief that the physical world is fundamentally mathematical in nature, then fundamental physics should not be cashed out in information-theoretic terms. Thus, the entropy of a black hole, if it is one, should not be identified with Shannon's information-theoretic entropy, and, a fortiori, neither should the surface area of a black hole's event horizon. Thus, proposition (v) is false.

Let us summarize the fate of Bekenstein's attempt to endow the analogy between the area of a black hole's event horizon and thermodynamic entropy as it appears in the Second Law with physical salience in the light of these considerations. If in my reconstruction of Bekenstein's argument towards the end of Section \ref{sec:argument} is correct, and if the above considerations in fact suffice to put in doubt propositions (ii), (iv), and particularly (vi), and to deny (v), then the propositions (vii) through (ix) cannot be maintained, at least not on the basis of this argument: we can no longer conclude that the horizon area of a black hole and thermodynamic entropy are essentially the same kind of physical quantity and that, therefore, black holes have entropies which contribute to the universe's total entropy as captured by the Generalized Second Law. Bekenstein's attempt to give physical salience to the formal analogy between the horizon area of black holes and thermodynamic entropy via Shannon's information-theoretic entropy as depicted in Figure \ref{fig:argstructure} does not succeed. 

Before we arrive at the conclusion, let me emphasize that the failure of a particular argument to a particular conclusion does not in itself establish the falsity of the conclusion: even though Bekenstein's original argument may fail to prove that black holes have entropies, this failure, importantly, does not imply that black holes are not thermodynamic objects. This global conclusion would only be warranted if {\em all} available arguments would fare equally badly.

\section{Conclusion}\label{sec:conc}

As attempts to formulate a quantum theory of gravity are bedevilled by a frustrating lack of empirical data to guide theoretical work, black hole thermodynamics in general, and the Bekenstein-Hawking formula for the black hole entropy in particular, are widely accepted as physical fact. This is so even though there is so far no empirical confirmation of any of the thermodynamic aspects of black holes.\footnote{But see Th\'ebault (this volume) for an analysis of whether analogue models can bestow confirmation on black hole radiation. Cf.\ also \citet{unr14} for an argument that Hawking radiation has in fact been measured (though indirectly).} This almost universal sentiment is, among many others, expressed by Jonathan Oppenheim is his obituary of Jacob Bekenstein:
\begin{quote}
Any potential theory of quantum gravity should correctly predict the value of the black hole entropy. Indeed, as quantum gravity is not driven by experiment, {\em black hole thermodynamics is the only really solid piece of information we have in our attempts to construct it.} \citep[805, emphasis added]{opp15}
\end{quote}

Why, if there is no empirical confirmation of the thermodynamic behaviour of black holes available, is entropy so generally predicated of black holes and why do virtually all physicists---in quantum gravity and elsewhere---accept the Bekenstein-Hawking formula as correctly capturing its quantity? Since there is no direct empirical confirmation, the reasons physicists have for accepting black hole entropy and radiation must derive from theoretical considerations. If my argument above is sound, then the original argument by Bekenstein with its detour through information theory does not succeed in establishing the physical salience of the otherwise merely formal analogy between thermodynamic entropy and the black hole area, and so cannot offer the basis for accepting black hole thermodynamics as ``the only really solid piece of information''. 

It may appear unfair, or even uninteresting, to attack an argument that could not, and did not, stand by itself: Bekenstein's papers making the case that black holes have entropy, and that furthermore, this entropy is measured by the event horizon's area, were not taken very seriously by physicists before Hawking's semi-classical calculation of his eponymous radiation \citep[\S5.2]{cur17a}. Rafael \citet{sor98} expresses this sentiment when he writes that the ``best known piece of evidence'' for the association between horizon area with entropy is Hawking radiation. As noted before, at the purely classical level, the thermodynamic analogy remains weak.\footnote{Though \citet{cur15} demurs.} For classical black holes, the obvious disanalogy is that since they do not radiate, they should not be ascribed a temperature, and so they are not thermodynamic objects. Although Bekenstein's formulae were already quantum in the sense that the definition of $S$ contained an $\hbar$, or, alternatively, the Planck length, this quantum trace originated simply from a classically completely free choice of a fiducial area for the event horizon, which might not have included any $\hbar$. While it is thus arguably the case that Bekenstein's argument did not stand by itself in the first place, I maintain that given the lack of direct empirical confirmation of any thermodynamic aspects of black holes a careful analysis of all major arguments remains an important source of foundational insight.

As a final point, it should be remarked that even if black holes have entropy, they may lack microstates specified by degrees of freedom which are qualitatively distinct from the macroscopic degrees of freedom. Classically, black holes have `no hair', i.e., the spacetime external to a black hole is uniquely determined by the black hole's mass, angular momentum, and charge. Thus, as far as classical GR in concerned, a black hole is fully characterized by its macrostate in terms of these three properties. In itself, it does not imply that it also has microstates defined by additional properties. As stated earlier, the general expectation is of course that in a quantum theory of gravity, black holes will be shown to have microstates in some sense and the way they get to have the entropy they have is explained by counting the microstates in the appropriate way. On such a Boltzmannian view, one would expect the entropy to be huge, since presumably very many distinct microstates would give rise to one and the same simple macrostate.

However, as \citet{chieal14} have argued, black holes may well have entropy while lacking microstates characterized by distinct---`hidden'---degrees of freedom. The entropy may be an entanglement entropy and arise from a coarse graining in the form of a split of the system into two components and the resulting restriction to the observables of one of the two subsystems.\footnote{The idea that the black hole entropy is an entanglement entropy resulting from tracing out degrees of freedom hidden behind the event horizon goes back to \citet{bomeal86}.} It is clear, however, that this line of reasoning requires the thermodynamic entropy of a black hole to be identified with the conceptually distinct von Neumann entropy, a claim which requires substantive defence. 

To return to the main topic of this article and to conclude, let me emphasize once again that even though Bekenstein's detour through information does not succeed in endowing the formal analogy between thermodynamic entropy and black hole horizon area with physical salience, it does not thereby follow that black holes are {\em not} thermodynamic objects. For instance, gedanken experiments concerning the limits of the amount of thermodynamic work that can or cannot be extracted from black holes lend some support to the idea that black holes are thermodynamic in nature.\footnote{I am thinking, in particular, of thought experiments unclouded by statistical and informational notions showing the consistent and beneficial application of thermodynamical concepts to black holes, such as most recently in \citet{prutim17}.} Moreover, the (direct or indirect) empirical confirmation of Hawking radiation would strongly support the idea, though of course in the case of an indirect detection questions regarding analogical reasoning would arise. Ultimately, as with any substantive claim in empirical science, only the usual kind of experimental and observational work can establish that black holes are thermodynamic objects.

\bibliographystyle{plainnat}
\bibliography{}

\end{document}